\documentclass[12pt]{article}

\begin{document}
\pagestyle{empty}
\begin{center}
{\Large \bf Conformal invariance and the expressions for       
$C_F^4\alpha_s^4$ contributions to the  
Bjorken polarized  and   Gross-Llewellyn Smith sum rules}

\vspace{0.1cm}

{A.L. Kataev }\\
\vspace{0.1cm}

Institute for Nuclear Research of the Academy of Sciences of
Russia,\\ 117312, Moscow, Russia\\
\end{center}
\begin{center}
{\bf ABSTRACT}
\end{center}
\noindent
Considering 
massless  axial-vector-vector  triangle 
diagram in the conformal invariant limit    
and  the results of the  analytical distinguished
calculations of the 5-loop  single-fermion circle 
corrections  to the QED $\beta$-function, we derive the analytical 
expressions for the  $C_F^4\alpha_s^4$-contributions
to  the Bjorken polarized
 and Gross-Llewellyn Smith sum rule. 
This visible in future evaluation can shed extra light on the 
reliability of the appearance of $\zeta_3$-term in the 
explicitly known part  of  5-loop corrections to the QED $\beta$-function 
which are proportional to the   conformal invariant set of 
   $C_F^4\alpha_s^4$ -contributions into  to  the $e^+e^-$-
annihilation Adler function. 
 \vspace*{0.5cm}

\pagestyle{plain}

In this article,  which is more detailed version of the  talk 
at Quarks-2008 International Seminar, presented 
prior the presentation at the same Seminar    
of the results of calculations of the 
complicated  analytical expression for   
the  non-singlet order $\alpha_s^4$  contribution  to the 
$e^+e^-$ annihilation Adler function function   
\begin{equation}
\label{adlerNS}
D^{NS}(Q^2)=Q^2\int_0^{\infty}
\frac{R(s)}{(s+Q^2)^2}ds
=3\sum_F Q_F^2C_D^{NS}(a_s(Q^2))
=3\sum_F Q_F^2
\bigg[1+\sum_{n=1}^{n=4}
d_n^{NS}a_s^{n}\bigg]
\end{equation}
\cite{Baikov:2008jh} 
we will try to clarify why to our point 
of view it is interesting and important to perform independent 
calculations of the $C_F^4\alpha_s^4$ contributions
to  the Bjorken polarized or  Gross-Llewellyn Smith sum rules 
of deep-inelastic lepton-nucleon scattering. In brief, these arguments 
were published  in Ref.\cite{Kataev:2008sk}.

In Eq.(1) 
$R(s)$   is the well-known $e^+e^-$ ratio, $Q_F$ are the quarks
charges,  $a_s=\alpha_s(Q^2)/\pi$ and  $\alpha_s(Q^2)$ is the 
$\rm{\overline{MS}}$-scheme QCD coupling constant, which obeys 
the property of asymptotic freedom  at large $Q^2$.
Note, that in the theoretical part of Eq.(1) the contributions
 of singlet-type diagrams were omitted,  in view 
of the fact that at the 5-loop level they are not yet  calculated,
but presumably and hopefully will be small.      

The calculation  of  $d_4^{NS}$  \cite{Baikov:2008jh} is    the third 
step after {\bf complete}  
analytical calculations of the $\alpha_s^2$ \cite{Chetyrkin:1979bj}
and $\alpha_s^3$ corrections \cite{Gorishnii:1990vf},
\cite{Surguladze:1990tg}
to the Adler function of vector currents. 
The analytical result for the $\alpha_s^3$ coefficient, obtained 
in Refs.\cite{Gorishnii:1990vf}, \cite{Surguladze:1990tg} was 
confirmed later on in  Ref. \cite{Chetyrkin:1996ez}. 
Its $SU(N)$-group structure 
was analyzed  in detail in Ref.  \cite{Broadhurst:1993ru}.
Some special features found in this work served the first 
theoretical argument in favor of the correctness of 
the results of Ref.\cite{Gorishnii:1990vf}.
    
Unfortunately, due to technical reasons (related to the desire 
to minimize  the huge  time of the complicated computer-based  calculations), 
the expression for 
the $\alpha_s^4$ contribution to Eq.(\ref{adlerNS}) 
 was presented  in the case of $SU(3)$ 
group only,  without singling out the corresponding Casimir operators
$C_F$ and $C_A$ \cite{Baikov:2008jh}. 
 This does not allow one  to study possible 
special theoretical  features of both   $\alpha_s^4$-coefficient to 
 $D^{NS}(Q^2)$ and to the photon vacuum polarization  constant $Z_{ph}$ 
in particular. The latter consideration was done at the $\alpha_s^3$-level 
in Ref. \cite{Gorishnii:1990vf}, where the observation was made   
that in the case 
of  $C_A=C_F=Tf/2=N$, which corresponds to the 
case of $SU(4)$ Supersymmetric Yang-Mills  (SYM) theory, first studied  
at three-loop order  
in Ref. \cite{Avdeev:1980bh}, 
$\zeta_3$-term, which appear  in the 4-loop approximation for $Z_{ph}$ 
in QCD, is canceling out. 
This observation gave the authors of Ref. \cite{Gorishnii:1990vf}
some additional confidence in favor of the validity of 
the results obtained. It will be highly desirable to understand whether 
the similar features are manifesting themselves at the 5-loop level.
  
Another 
interesting part of analytical result of
Ref.\cite{Baikov:2008jh}, namely the the  five-loop single-fermion    
contribution to the QED $\beta$-function, which is proportional 
to the single-fermion QED contribution to Eq.(\ref{adlerNS}),  
was  already  presented  
in the literature some time ago  \cite{Baikov:2007}. It has the 
following form
\begin{eqnarray}       
\label{beta{QED}}
\beta_{QED}^{[1]}&=&\frac{4}{3}{\rm{A}}+4{\rm A}^2-2{\rm A^3}-46{\rm{A^4}}
+\bigg(\frac{4157}{6}+128\zeta_3\bigg){\rm A}^5 \\ \nonumber
&=&\frac{4}{3}{\rm{A}}\times C_D^{NS}(\rm{A})
\end{eqnarray}
where ${\rm A}=\alpha/(4\pi)$ and $\alpha$ is the QED coupling constant,
which does not depend on any scale.
Indeed,  the coefficients of Eq.(\ref{beta{QED}}) 
are scheme-independent, at least in the schemes, not related 
to the  lattice regularization  \cite{Broadhurst:1992za}. 
The performed by different methods     
calculations  of the order $O({\rm A}^4)$-approximation  to  
Eq.(\ref{beta{QED}})~ \cite{Gorishnii:1990kd},~\cite{Broadhurst:1999zi} 
are supporting  this feature.

It is interesting, that   
the analytical structure of the 5-loop   result of Ref. 
\cite{Baikov:2007} {\bf differs} from the structure of the  previously known terms:
it contains  $\zeta_3$-term in the presented 5-loop  coefficient.  

Indeed,   at the intermediate stages of calculations of the 3-loop 
correction  to Eq.(\ref{beta{QED}}), which were done in Refs. 
\cite{Rosner}, \cite{Bender:1976pw} for the purpose 
of more detailed study 
of  the ``finite QED program''  \cite{Baker:1969an},  
$\zeta_3$-terms  were  appearing,  but they  canceled out  in the ultimate 
result.
Moreover, in Ref. \cite{Bender:1976pw} this feature was  
related to the  property of the conformal invariance of 
this part of QED $\beta$-function,     
though no proofs or references were given.
Note, that the calculations of Ref.\cite{Rosner} were done within the context 
of regularization with upper cutt-off in the momentum space. 
The results of these calculations were confirmed in the dimensional 
regularization  in 
the unpublished preprint of Ref.\cite{Chetyrkin:1980sa}. 
In this work the manifestation of the  feature of cancellation 
of $\zeta_3$-term in the 3-loop calculations 
was clearly  demonstrated at  the diagrammatic 
language. 
 
Next, in the process of evaluation  of the  4-loop term in 
Eq.(\ref{beta{QED}})  \cite{Gorishnii:1990kd}
the contributions with two  transcendentalities 
$\zeta_3$ and $\zeta_5$ appeared  at the intermediate stages 
of calculations, but these contributions canceled   in the 
final result. 
This feature was also expected \cite{Bender:1988nt}.
A knot-theory explanation of the cancellation of 
transcendentalities in the discussed above 3-loop and 4-loop 
QED results was given in Ref. \cite{Broadhurst:1995dq}.
Moreover, the strong  statement that the complete cancellation of 
transcendentalities from this part of QED $\beta$-function
may be expected at every order, was made by the authors 
of Ref.\cite{Broadhurst:1995dq}.
 Thus,  at  the 5-loop level one may expect, that 
$\zeta_3$, $\zeta_5$ and $\zeta_7$ should appear at the intermediate 
stages of concrete calculations,  
but cancel down in the final result.
However, Eq.(2) demonstrate that for $\zeta_5$ and $\zeta_7$ 
this property is valid, while for $\zeta_3$ this is not the case!

I do not know
whether this observation  may be related to the unproved property   
of ``maximal  transcendentality'', which at present is widely discussed 
while considering perturbative series for different quantities   
in the conformal invariant  $N=4$ 
SYM theory 
(see e.g. \cite{Kotikov:2006ts}, \cite{Drummond:2007aua}).
Thus we do not know whether the appearance of the transcendental 
term  may be considered {\it pro or contra} the validity of the 
 results of Refs. \cite{Baikov:2007}, \cite{Baikov:2008jh}.
In any case, to clarify the status of this new feature 
of perturbative series in QED it is highly desirable  to perform 
{\it independent calculational studies}  of 
the  results of Ref.  \cite{Baikov:2007},  \cite{Baikov:2008jh}.

Let me   discuss in more detail one of the possibilities for   
performing  some clarifications of the status  of the results of Ref.\cite{Baikov:2007}
(for brief presentation  see Ref.\cite{Kataev:2008sk}). 
It is based on the property of the conformal symmetry 
of the sets of the diagrams considered and 
the relation, 
derived by   Crewther in Ref.  \cite{Crewther:1972kn} using 
the concept of conformal symmetry, typical to  quark-parton
model of strong interactions.   
The  consequences of this relation were studied     
in Ref. \cite{Adler:1973kz} and Ref.~\cite{Broadhurst:1993ru}
in the conformal-invariant limit of QCD  and were generalized by 
different ways to the case of full QCD with non-zero QCD   
$\beta$-function in the works of Ref.\cite{Broadhurst:1993ru}, 
\cite{Gabadadze:1995ei},\cite{Brodsky:1995tb},\cite{Kataev:1996ce},
\cite{Rathsman:1996jk},\cite{Crewther:1997ux} 
(for a review see \cite{Braun:2003rp}).

We will follow here the work of Ref.\cite{Gabadadze:1995ei}. 
Let us translate  the investigations   of Refs. \cite{Crewther:1972kn}, 
\cite{Adler:1973kz}, performed in the $x$-space, to   the language 
of the momentum space following the presentations, given in
 Ref.\cite{Gabadadze:1995ei} (some additional 
details can  be found  and in Ref. \cite{Kataev:1996ce}). 

Consider first the axial-vector-vector (AVV)  3-point function
\begin{equation}
T_{\mu\alpha\beta}^{abc}(p,q)=i\int<0|TA_{\mu}^{a}(y)V_{\alpha}^{b}(x)
V_{\beta}^{c}(0)|0>e^{ipx+iqy}dxdy=d^{abc}T_{\mu\alpha\beta}(p,q)
\label{1}
\end{equation}
where
$A_{\mu}^{a}(x)=\overline{\psi}\gamma_{\mu}\gamma_{5}(\lambda^{a}/2)\psi$,
$V_{\mu}^{a}(x)=\overline{\psi}\gamma_{\mu}(\lambda^{a}/2)\psi$
are the axial and vector  non-singlet quark currents. The r.h.s.
of Eq.(\ref{1}) can be expanded  in a basis of 3 independent tensor
structures under the condition $(pq)=0$ as
\begin{equation}
\begin{array}{rcl}
T_{\mu\alpha\beta}(p,q) & = &
\xi_1(q^2,p^2)\epsilon_{\mu\alpha\beta\tau}p^{\tau}\\
 & +  &
\xi_2(p^2,q^2)(
q_{\alpha}\epsilon_{\mu\beta\rho\tau}p^{\rho}q^{\tau}-
q_{\beta}\epsilon_{\mu\alpha\rho\tau}p^{\rho}q^{\tau})\\
 & +  &
\xi_3(p^2,q^2)(
p_{\alpha}\epsilon_{\mu\beta\rho\tau}p^{\rho}q^{\tau}+
p_{\beta}\epsilon_{\mu\alpha\rho\tau}p^{\rho}q^{\tau})~~~.
\end{array}
\label{2}
\end{equation}
Taking now the divergency of axial current one can get the following
relation for the invariant amplitude $\xi_1(q^2,p^2)$:
\begin{equation}
q_{\beta}T_{\mu\alpha\beta}(p,q)=\epsilon_{\mu\alpha\rho\tau}
q^{\rho}p^{\tau}\xi_1(q^2,p^2)
\label{3}
\end{equation}
while the property of the conservation of the vector currents implies
that
\begin{equation}
lim|_{p^2\rightarrow{\infty}}p^2\xi_3(q^2,p^2)=-\xi_{1}(q^2,p^2)
\label{4}
\end{equation}
(see Ref.\cite{Gabadadze:1993uc} for the discussions of 
the details of the derivation
of Eqs.(\ref{2})-(\ref{4})).

In order to clarify the meaning of the second invariant amplitude
$\xi_2(q^2,p^2)$, let us first define the characteristics of
the deep-inelastic processes, namely the polarized Bjorken sum rule
\begin{equation}
Bjp(Q^2)=\int_0^1[g_1^{ep}(x,Q^2)-g_1^{en}(x,Q^2)]dx=\frac{1}{6}
g_AC_{Bjp}(a_s)=\frac{1}{6}g_A\bigg[1+\sum_{n=1}^{n=4}
c_n a_s^n\bigg]
\label{5}
\end{equation}
and the Gross-Llewellyn Smith sum rule
\begin{equation}
GLS(Q^2)=\frac{1}{2}\int_0^{1}\bigg[F_3^{\nu p}(x,Q^2)
+F_3^{\overline{\nu}p}(x,Q^2)\bigg]dx= 3C_{GLS}(a_s)~~~~.  \label{6}
\end{equation} 
where $a_s=\alpha_s/\pi$. 
The coefficient function $C_{Bjp}(a_s)$ can be found
from the operator-product expansion of two non-singlet vector
currents ~\cite{Gorishnii:1985xm}
\begin{equation}
i\int T{V_{\alpha}^{a}(x)V_{\beta}^{b}(0)}e^{ipx}dx|_{p^2\rightarrow{\infty}}
\approx C_{\alpha\beta\rho}^{P,abc}A_{\rho}^{c}(0)+other~structures
\label{7}
\end{equation}
with 
\begin{equation}
C_{\alpha\beta\rho}^{P,abc}\sim
id^{abc}\epsilon_{\alpha\beta\rho\sigma}
\frac{p^{\sigma}}{P^2}C_{Bjp}(a_s)~~~.  \label{8} \end{equation}
and $P^2=-p^2$.
In the case of the definition of the coefficient function of the
Gross-Llewellyn Smith sum rule one should consider the
operator-product expansion of the axial and vector  non-singlet
currents
\begin{equation}
i\int T{A_{\mu}^{a}(x)V_{\nu}^{b}(0)}e^{iqx}dx|_{q^2\rightarrow{\infty}}
\approx C_{\mu\nu\alpha}^{V,ab}V_{\alpha}(0)+other~structures~~~
\label{9}
\end{equation}
where  
\begin{equation}
C_{\mu\nu\alpha}^{V,ab}\sim
i\delta^{ab}\epsilon_{\mu\nu\alpha\beta}
\frac{q^{\beta}}{Q^2}C_{GLS}(a_s)~~~
\label{10}
\end{equation}
and $Q^2=-q^2$.
The third important quantity, which will enter into our analysis, is
the QCD coefficient function $C_D^{NS}(a_s)$ of the Adler $D$-function
of the  non-singlet axial currents
\begin{equation}
D^{NS}(a_s)=-12\pi^2q^2\frac{d}{dq^2}\Pi_{NS}(q^2)
=3\sum_F Q_F^2C_D^{NS}(a_s)
\label{11}
\end{equation}
with  $\Pi_{NS}(q^2)$  defined as
\begin{equation}
i\int<0|TA_{\mu}^a(x)A_{\nu}^b(0)|0>e^{iqx}dx=\delta^{ab}(g_{\mu\nu}q^2-
q_{\mu}q_{\nu})\Pi_{NS}(q^2)~~~.
\label{12}
\end{equation}

At this point we will stop with definitions of the basic quantities
and return to the consideration of the 3-point function of Eq.(\ref{1}).
Following original work \cite{Crewther:1972kn} one can apply to 
this correlation function an
operator-product expansion in the limit $|p^2|>>|q^2|$,
$p^2\rightarrow \infty$, namely expand first the $T$-product of two
non-singlet vector currents via Eq.(\ref{7}) and then take the vacuum
expectation value of the $T$-product of two remaining non-singlet
axial currents defined through Eq.(\ref{12}). 
These studies imply that \cite{Gabadadze:1995ei} 
\begin{equation}
\xi_2(q^2,p^2)|_{|p^2|\rightarrow\infty}\rightarrow \frac{1}{p^2}
C_{Bjp}(a_s)\Pi_{NS}(a_s)
\label{13}
\end{equation}
and thus
\begin{equation}
q^2\frac{d}{dq^2}\xi_2(q^2,p^2)|_{|p^2|\rightarrow\infty}\rightarrow
\frac{1}{p^2}C_{Bjp}(a_s)C_D^{NS}(a_s)~~~.
\label{14}
\end{equation}
Equations (\ref{13}),(\ref{14}) reflect the physical meaning of the invariant
amplitude $\xi_2(q^2,p^2)$ and should be considered together with the
relations for the invariant amplitudes $\xi_1(q^2,p^2)$ of  Eq.(\ref{3})
and $\xi_3(q^2,p^2)$ from  Eq.(\ref{4})).

On the other hand, it was shown in Ref.\cite{schreier} that in a conformal
invariant (c-i)  limit the three-index tensor of Eq.(\ref{1}) is proportional
to the fermion triangle one-loop graph, constructed from the
massless fermions: 
\begin{equation}
T_{\mu\alpha\beta}^{abc}(p,q)|_{c-i}=
d^{abc}K(a_s)\Delta_{\mu\alpha\beta}^{1-loop}(p,q)~~~.
\label{15}
\end{equation}
In other words, in a conformal invariant limit one has
\begin{eqnarray}
\label{16}
\xi_1^{c-i}(q^2,p^2) & = & K(a_s)\xi_1^{1-loop}(q^2,p^2)~~~,\\
\xi_2^{c-i}(q^2,p^2) & = & K(a_s)\xi_2^{1-loop}(q^2,p^2)~~~,\\
\xi_3^{c-i}(q^2,p^2) & = & K(a_s)\xi_3^{1-loop}(q^2,p^2)~~~.
\end{eqnarray}
Moreover, in view of the Adler-Bardeen theorem~\cite{Adler:1969er}
the invariant amplitude $\xi_1(q^2,p^2)$, related to the
divergency of axial current (see Eq.(\ref{3})), has no radiative corrections.
Therefore 
one has $K(a_s)=1$.
The 3-loop light-by-light-type scattering graphs, which were
calculated in Ref.  \cite{Anselm:1989gi} and analyzed in
Ref. \cite{Efremov:1989ry}  , do not affect this conclusion. 
Indeed, in the case
of the 3-point function of the  non-singlet
axial-vector-vector currents they are contributing to the
higher order QED corrections, while the QCD corrections of the
similar origin are appearing only in the 3-point function
with the  singlet axial current in one of the vertexes,
which will be not discussed here.

Taking into account the property 
$K(a_s)=1$
for the   3-point function of the
non-singlet axial-vector-vector currents allows us 
to derive the fundamental Crewther relation
\begin{equation}
C_{Bjp}(a_s(Q^2))C_D^{NS}(a_s(Q^2))|_{c-i}=1~~~,
\label{17}
\end{equation}
which is  valid in the conformal
invariant limit in all orders of perturbation theory. The similar
relation is also true for the  coefficient function $C_{GLS}(a_s)$,
defined by  Eqs.(\ref{6}),~(\ref{9}),~(\ref{10}) ~\cite{Adler:1973kz} . 
Indeed, considering first
the  operator-product expansion of the axial and vector  non-singlet
currents 
 in the 3-point function of Eq.(\ref{1})  (see Eq.(\ref{9}) and 
Eq.(\ref{10})), 
taking  the $T$-product of the remaining vector currents and
repeating the above   discussed  analysis, one can find that in the
conformal invariant  limit the second  identity takes place:
\begin{equation}
C_{GLS}(a_s(Q^2))C_D^{V}(a_s(Q^2))|_{c-i}=1~~~
\label{18}
\end{equation}
where $C_D^{V}(a_s)$ is the coefficient function of the Adler
$D$-function of two vector currents, which coincide 
with $C_D^{NS}(a_s(Q^2))$ when the high-order 
 singlet light-by-light type  contributions
to $C_D^{V}$,  
which appear first at the $\alpha_s^3$-level \cite{Gorishnii:1990vf}, 
are not yet taken into account  (see Ref.\cite{Baikov:2008jh}).

The results of calculations of Ref.\cite{Baikov:2007}  
are equivalent to the following 
expression for the $C_D^{NS}(a_s)$ function
\begin{equation}
\label{Baikov2007}
C_D^{NS}(a_s)=\bigg[1+\frac{3}{4}{\rm C_F}a_s-\frac{3}{32}
{\rm C_F^2}a_s^2-\frac{69}{128}{\rm C_F^3}a_s^3+\bigg(\frac{4157}{2048}
+\frac{3}{8}\zeta_3\bigg){\rm C_F^4}a_s^4 \bigg]
\end{equation}
where ${\rm C_F}=(N^2-1)/(2N)$. Using now Eq.(\ref{17}) and Eq.(\ref{18}) we get the new 
analogous scheme-independent  contributions 
to the Bjorken polarized sum rule 
and Gross-Llewellyn Smith sum rule
\begin{equation}
\label{results}
C_{Bjp}(a_s)=C_{GLS}(a_s)=1-\frac{3}{4}{\rm C_F}a_s
+\frac{21}{32}{\rm C_F^2}a_s^2 -\frac{3}{128}{\rm C_F^3}a_s^3
-\bigg(\frac{4823}{2048}+\frac{3}{8}\zeta_3\bigg){\rm C_F^4}a_s^4
\end{equation}
Note, that the order $a_s$, $a_s^2$ and $a_s^3$ coefficients 
are in agreement with the result of explicit calculations, performed 
in Refs.\cite{Kodaira:1979ib},~\cite{Gorishnii:1985xm} and~ \cite{Larin:tj}
respectively. Thus, the direct   calculation of the predicted $a_s^4$
coefficient  may be rather useful for the  independent 
cross-checks of the results of Ref.\cite{Baikov:2007}  
and the verification of the appearance 
of $\zeta_3$-term at the 5-loop level.

These calculations  may 
give us the hint whether the appearance of   
of $\zeta_3$-term in the result  of Eq.(\ref{Baikov2007})
{\it is the new mathematical  feature}, which appear 
in higher orders of perturbation theory.

If results of  possible directcalculations of the contributions 
to the Bjorken sum rule  will 
agree with the presented expression, then the appearance of 
$\zeta_3$-term in the 5-loop correction to the QED 
$\beta$-function and in  the $C_F^4\alpha_s^4$ contribution 
into  the $e^+e^-$ annihilation Adler function may get 
independent support and may be 
analyzed  within the framework of the recently introduced 
concept of ``maximal 
transcendentality''. 

After the possible cross-check of the results of 5-loop direct calculations 
it may be of interest to study why the  
the Lipatov-type estimates  
originally proposed in Ref. \cite{Lipatov:1976ny}
for the sign-alternating series in the $g\Phi^4$-theory
(for a review see Ref. \cite{Kazakov:1980rd}) 
are not working properly in QED. Indeed,  theoretical works 
of   Refs.\cite{Itzykson:1977mf},  \cite{Balian:1977mk} indicate, 
that the QED series for $\beta_{QED}^{[1]}$-function should also   
have sign-alternating behavior, in contradiction 
to the concrete existing results  of Ref.\cite{Gorishnii:1990kd} and  
Ref. \cite{Baikov:2007} (see Eq.(\ref{beta{QED}})).

{\bf Acknowledgments}

I wish to thank P.A. Baikov, K.G. Chetyrkin and J.H.  Kuhn  for 
discussions and private communications of the problems, related to their  intriguing 
5-loop project.
I am grateful to  D. J. Broadhurst and G. T. Gabadadze for 
pleasant  collaboration and useful  discussions of various 
scientific problems, related to single-scale generalizations 
of the Crewther studies. It is the pleasure to thank  G.P. Korchemsky for  
useful comments concerning the current situation 
in $N=4$ SYM theory and 
 J.L. Rosner
for his interest in the results, published in Ref.\cite{Kataev:2008sk},
useful questions and pointing out the existence of the 
knot-theory study of Ref.\cite{Broadhurst:1995dq}.
The important  contributions to 
to the  long-standed   calculational  research   of all  those,  
closely  related to Physics Department 
of Moscow States University, which is celebrating   
70-th anniversary in October 2008,   is acknowledged in 
the concrete   references below.  
This work is done within the framework of RFBR Grants 
Grants  08-01-00686-a and 06-02-16659-a.

\end{document}